\documentclass[letterpaper,10pt,conference,dvipsnames]{ieeeconf} \IEEEoverridecommandlockouts                          
\usepackage{graphicx}
\usepackage{amsmath}
\usepackage{amssymb}
\usepackage{mathtools}
\usepackage{xcolor}
\usepackage{theorem}
\usepackage{tikz}
\usepackage{pgfplots}
\usepackage{pgf}
\usepackage{cite}
\usepgfplotslibrary{colorbrewer}
\usepackage[font=footnotesize,skip=0.1pt]{caption}
\usepackage{todonotes}

\DeclareMathOperator{\E}{\mathbb{E}}
\DeclareMathOperator{\Z}{\mathbb{Z}}
\DeclareMathOperator{\F}{\mathcal{F}}

\newtheorem{lemma}{Lemma}
\newtheorem{remark}{Remark}
\newtheorem{theorem}{Theorem}

\newcommand*\circled[1]{\tikz[baseline=(char.base)]{
            \node[shape=circle,draw,inner sep=0.8pt] (char) {\scriptsize #1};}}

\title{\LARGE \bf Resource-Aware Stochastic Self-Triggered Model Predictive Control}

\author{Yingzhao Lian, Yuning Jiang, Naomi Stricker, Lothar Thiele, Colin N. Jones% <-this % stops a space
\thanks{This work has received support from the Swiss National Science Foundation under the RISK project (Risk Aware Data-Driven Demand Response), grant number 200021 175627, and from the Swiss National Science Foundation under the NCCR Automation project, grant agreement 51NF40\_180545.}
\thanks{Yingzhao Lian, Yuning Jiang and Colin N. Jones are with Automatic Control Laboratory, EPFL, Switzerland.{\tt$\{$yingzhao.lian, yuning.jiang,colin.jones$\}$@epfl.ch}}
\thanks{Naomi Stricker and Lothar Thiele are with Computer Engineering and Networks Laboratory, ETH Zürich, Switzerland.{\tt $\{$nstricker, thiele$\}$@ethz.ch}}}

\begin{document}
\maketitle
\thispagestyle{empty}
\pagestyle{empty}
%%%%%%%%%%%%%%%%%%%%%%%%%%%%%%%%%%%%%%%%%%%%%%%%%%%%%%%%%%%%%%%%%%%%%%%%%%%%%%%%
\begin{abstract}
This paper considers the control of uncertain systems that are operated under limited resource factors, such as battery life or hardware longevity. We consider here resource-aware self-triggered control techniques that schedule system operation non-uniformly in time in order to balance performance against resource consumption. 

When running in an uncertain environment, unknown disturbances may deteriorate system performance by acting adversarially against the planned event triggering schedule. In this work, we propose a resource-aware stochastic predictive control scheme to tackle this challenge, where a novel zero-order hold feedback control scheme is proposed to accommodate a time-inhomogeneous predictive control update.
\end{abstract}

%%%%%%%%%%%%%%%%%%%%%%%%%%%%%%%%%%%%%%%%%%%%%%%%%%%%%%%%%%%%%%%%%%%%%%%%%%%%%%%%
\section{INTRODUCTION}
\label{sec::Intro}Most devices in Internet of Things (IoT) networks and wireless sensing systems are operated with some limited resource factors, such as battery life or hardware longevity. In order to maintain desirable performance, a minimal number of triggers are required to best exploit the limited resource. Event-triggered control and self-triggered control are two main control schemes~\cite{heemels2012introduction} accommodating this issue. In particular, control under an event-triggered scheme is updated \emph{reactively} by determining a trigger condition, for which a sensor has to continuously monitor the trigger condition. Contrarily, a self-triggered scheme updates \emph{proactively} by planning the next trigger in advance, leaving the sensor and controller in idle mode. Due to the limitation of the resource factors, especially battery life, a self-triggered scheme is preferable and is, therefore, the research object of this work.

The key ingredient of a self-triggered controller is the decision of the triggering time sequence. The triggering time can be chosen as long as possible to minimize resource consumption as in~\cite{berglind2012self, bernardini2012energy}. However, to balance performance and resource consumption more effectively, the response of the resource is explicitly considered in the model predictive control (MPC) problem in~\cite{henriksson2012self,wildhagen2019resource}. The former work solves a mixed-integer problem and is designed for discrete-time systems, while the latter work solves a non-convex continuous-time optimal control problem, and has been later generalized to a distributed control scheme~\cite{lian2020resource}. 

Running a triggered system within an uncertain environment while maintaining system performance is challenging. Especially for the self-triggered controllers, the lack of sensor measurement between consecutive triggers requires extra consideration of the uncertainty propagation. In~\cite{li2014event}, a nominal control law is determined based on a nominal system, while the discrepancy between the nominal and measured trajectories serves as the triggering condition. In~\cite{aydiner2015robust,brunner2016robust,farina2012tube}, the idea of tube-MPC enables the design of robust self-triggered controllers for both discrete-time and continuous time linear systems. In~\cite{liu2018robust}, a min-max optimization is used to optimize the worst-case performance. Though it is capable of handling general uncertainties in nonlinear systems, the resulting non-convex robust optimization problem is NP-hard~\cite{ben2009robust}. Except for~\cite{liu2018robust}, other previous works mainly decouple the effects of uncertainty from the nominal system, and the feedback control laws are all updated with a fixed frequency.

In this work, a resource-aware stochastic predictive control scheme is designed for a stochastic linear system where the process noise is explicitly considered in the predictive control problem. In particular, a discrete-time zero-order-hold linear feedback control law is integrated into the closed-loop predictive control problem. The update time instances of this feedback control law distribute non-uniformly on the time axis, which we term \textit{time-inhomogeneous}, and are optimized within the predictive control problem. The contributions of this work are summarized into two aspects:
\begin{itemize}
    \item A sigma field decomposition strategy is proposed to enable the analysis of a time-inhomogeneous control.
    \item A discrete-time closed-loop feedback control law for stochastic self-triggered MPC is proposed.
\end{itemize}

The rest of this paper is organized as follows: Section~\ref{sec::review} reviews deterministic resource-aware self-triggered MPC, after which the stochastic extension is elaborated in Section~\ref{sec::method}. This section further details the sigma field decomposition and the continuous-time dynamics of a discrete-time feedback law. The effectiveness of the proposed method is validated in Section~\ref{sec::result} and conclusions are given in Section~\ref{sec::concl}.

\textbf{Notation:} $\{x_i\}_{i=0}^K$ denotes a finite set of size $K$ whose elements $x_i$ are indexed by $i$. $\Z_a^b$ is the set of integers $\{a,a+1\dots,b\}$. $A\backslash B:=\{x\,|\,x\in A,\, x\notin B\}$. $\E\{\cdot\}$ denotes the expectation operator and $\mathbb{P}(\cdot)$ represents the probability.

\section{Deterministic Self-Triggered MPC}
\label{sec::review}
This section recaps the main idea of deterministic resource-aware self-triggered control. We consider a deterministic continuous time LTI system:
\begin{align}\label{eq::dyn}
    \frac{dx(t)}{dt} = Ax(t)+Bu(t)
\end{align}
with state $x(\cdot):[0,\infty)\to\mathbb{R}^{n_x}$ and control input $u(\cdot):[0,\infty)\in\mathbb{R}^{n_u}$. A self-triggered controller determines both the value of the control inputs and the time instances at which the control input is changed. In the framework of direct optimal control~\cite{bock1984multiple}, a self-triggered controller parameterizes its control inputs over the time horizon $[0,t_N]$ by
\begin{align}\label{eqn:deter_ctrl_1}
    u(t) = \sum\limits_{k=0}^{N-1}v_k\zeta_k(t,t_k,t_{k+1})\;,
\end{align}
where $v\in\mathbb{R}^{Nn_u}:=[v_0^\top,v_1^\top\dots,v_{N-1}^\top]^\top $ denotes the stacked control coefficient vector and the orthogonal functions $\zeta_k\in\mathcal{L}^2[t_0,t_N],\;k\in\Z_0^{N-1}$ model the triggering property with a piece-wise constant function
\begin{align}\label{eqn:deter_ctrl_2}
    \zeta_k(t,t_k,t_{k+1})=\begin{cases} 1 & t\in(t_k,t_{k+1}]\\
    0& \text{otherwise}\;.
    \end{cases}
\end{align}
For the sake of compactness, we define the triggering time interval $\Delta_k:=t_{k+1}-t_k$ and use the notation $\Delta = [\Delta_0,...,\Delta_{N-1}]^\top$. A self-triggered agent updates its control inputs at triggering time instances $\{t_k\}_{k=0}^{N-1}$. When the control law is fixed within $(t_k,t_{k+1}]$, the resource $r$ is recharged at a constant rate $\rho$ until saturation. More specifically,
\begin{align*}
    \forall\,t\in [t_k,t_{k+1}),\quad \dot{r}(t) = \delta(\overline{r}-r(t))\rho\;,
\end{align*}
where $\overline{r}$ is the saturated value and $\delta(\cdot)$ is a step function with $\delta(s)=1$ if $s>0$ and 0 elsewhere. When the agent is triggered to update the control input, the resource is discharged by an amount of $\eta(\Delta_k)$ to pay the update cost. Hence, the resource at triggering times $\{t_k\}_{k=0}^{N-1}$ is
\begin{align}\label{eqn:resource_dyn}
r(t)=\begin{cases} r_0&\;\;t=t_0\\[0.16cm]
\lim\limits_{t\rightarrow t_k^-} r(t)-\eta(\Delta_k)&\;\; t\in\{t_k\}_{k=1}^{N-1}
\end{cases}
\end{align}
with an initially available resource  $r_0$ at $t_0$. Here, $t\to t_k^-$ represents the left limits, i.e., $t\to t_k$ and $t<t_k$. Moreover, the resource $r$ is required to be lower bounded by $\underline{r}$. In conclusion, a resource-aware self-triggered agent can update its control input when its resource is sufficiently high to stay above the lower bound $\underline{r}$. Otherwise, it must wait until enough resources are available. Once the controller is triggered at time $t_0$, the resource-aware self-triggered control solves the following optimization problem to plan the next trigger time $t_1$ and the control input within $[t_0,t_1]$,
\begin{subequations}\label{eqn:deter_resource_mpc}
\begin{align}
    \min_{x(\cdot),v,\Delta}\;\;& \sum\limits_{k=0}^{N-1}\int_{t_{k}}^{t_{k+1}} l(x(\tau),v_k)d\tau+ M(x(t_N))\label{eq::obj}\\
    \text{s.t.}\quad\; &\;x(t_0)=x_0\;,\;r(t_0) = r_0\label{eq::initial}\\
    &\forall\,t\in[t_0,t_N],\; \frac{dx(t)}{dt} = Ax(t)+Bu(t),\\
    &\forall\,t\in[t_0,t_N],\; x(t)\in\mathcal{X}\;,\;u(t)\in\mathcal{U}\;,\label{eq::stateCons}\\
    &\forall\,k\in\{0,1,...,N-1\}\notag\\
    &r(t_{k+1}) = \min\{\rho \Delta_k+r(t_k)-\eta(\Delta_k),\overline{r}\}\label{eqn:discrete_resource_1}\\
    & r(t_{k+1})\in[\underline{r},\overline{r}]\;,\label{eqn:resouce_cons}\\
    &\Delta_k\in[\underline{\Delta},\overline{\Delta}]\label{eqn:delta_cons}
\end{align}
\end{subequations}
where the control input $u(\cdot)$ is defined in~\eqref{eqn:deter_ctrl_1}, $l(\cdot,\cdot)$ and $M(\cdot)$ in~\eqref{eq::obj} are stage cost and terminal cost, respectively. ~\eqref{eqn:discrete_resource_1} is a simplified yet equivalent formulation of the resource dynamics~\eqref{eqn:resource_dyn} and the resource is bounded  by~\eqref{eqn:resouce_cons}. The constraints of the triggering time interval in~\eqref{eqn:delta_cons} protects the system from being Zeno or frozen. $\mathcal{X}\subseteq \mathbb{R}^{n_x}$ and $\mathcal{U}\subseteq\mathbb{R}^{n_u}$ in~\eqref{eq::stateCons} model the states and inputs constraints. The initial condition is given by~\eqref{eq::initial}.

\section{Stochastic Self-Triggered MPC}
\label{sec::method}
In this section, we consider the linear time invariant system~\eqref{eq::dyn} contaminated by a Wiener process noise. This is described by the stochastic differential equation (SDE)
\begin{align}\label{eqn:sde_lin}
    dx(t) = (Ax(t)+Bu(t))dt + dW\;,
\end{align}
where $W$ denotes a multi-dimensional Wiener process with statistics
\begin{align*}
    \E\{W(t)W(s)\}=Q\min(s,t)\;,\;\; \E\{W(t)\} = 0\;.
\end{align*}
The open-loop evolution of the system's state distribution~\eqref{eqn:sde_lin} is widely studied in filter theory~\cite{krishnan2013nonlinear} and the state evolution remains Gaussian $\mathcal{N}(\mu(t),\;P(t))$ where
\begin{subequations}\label{eqn:prob_dyn}
\begin{align}
    \frac{d\mu(t)}{dt} &= A\mu(t)+ Bu(t)\;,\label{eqn:prob_dyn_mu}\\
    \frac{dP(t)}{dt}&= AP(t)+P(t)A^\top+Q\;\label{eqn:prob_dyn_var}.
\end{align}
\end{subequations}
Above all, given the dynamics in~\eqref{eqn:prob_dyn}, it is trivial to adapt the deterministic formulation in~\eqref{eqn:deter_resource_mpc} to generate an open-loop resource-aware stochastic MPC. The focus and main contributions of this work are to develop a \textbf{closed-loop} scheme with respect to the dynamics~\eqref{eqn:prob_dyn}. In particular, a feedback control law is explicitly considered in the predictive control problem and this feedback control law should satisfy following requirements:
\begin{enumerate}
    \item The feedback control law can only change its value when the controller is triggered, otherwise, the control inputs remain constant.
    \item The feedback control law is not updated at a fixed frequency, and its update time instances are decision variables of the self-triggered problem. 
\end{enumerate}

In the following, the dynamics of the state distribution driven by a discrete time feedback control law are developed based on the sigma field decomposition technique. This dynamics results in a resource-aware stochastic self-triggered MPC, whose numerical implementation is discussed at the end. In order to convey the elegance of the proposed scheme, we state the main results intuitively in this section while the obscure math details are attached in the Appendix.

\subsection{Stochastic Process Decomposition}\label{subsec::decomp}
Considering an ordered triggering time sequence $\{t_k\}_{k=0}^{N}$, a sigma field $\F_k$ collects all the stochastic events occurring between $[t_0,t_k]$, particularly, $\F_0$ includes all the deterministic events. Because the controller can only update when it is triggered, we propose to partition the stochastic events by time intervals. In particular, the collection of stochastic events between two consecutive triggers is defined by $\F_{k,k+1}:=\sigma(\F_{k+1}\backslash\F_{k})$, where $\sigma(\cdot)$ denotes the minimal sigma field. The following lemma indicates that there is no information loss with the partitioning $\{\F_{k,k+1}\}_{k=0}^{N-1}$. This result will serve as the key component of the feedback control law analysis. 

\begin{lemma}\label{lem:decomp_sigma}
For a given Wiener process $W$ with a stopping time sequence $\{t_k\}_{k=0}^{N-1}$, if $t_j> t_i$ holds almost surely for all $j>i$, the sigma field at time $t_{N}$ can be decomposed as
$\F_{N} = \sigma(\cup_{i=0}^{N-1}\F_{i,i+1})$,
where $\F_{i,i+1}\perp\F_{j,j+1}$ holds for all $i\neq j$. 
\end{lemma}

The proof can be found in Appendix~\ref{sect:append}. Lemma~\ref{lem:decomp_sigma} enables us to decompose the statistics of the state evolution into non-overlapping time intervals. We have a special focus on the decomposition of the covariance matrix $P(t)$ because of its close link with the feedback control law. The projection of the covariance matrix $P(t)$ onto the stochastic events within $(t_k,t_{k+1}]$ is defined by $P_k(t):=\E(P(t)|\F_{k,k+1})$, and  Lemma~\ref{lem:decomp_sigma} implies that
\begin{align}\label{eqn:P_decomp}
    \forall\, t\in[t_0,t_{N}]\;,\;\;P(t) = \sum\limits_{i=0}^{N-1}P_i(t)\;.
\end{align} 
Based on this decomposition, the open-loop evolution of the conditional dynamics of $P_k(t)$ are given by  
\begin{align}\label{eqn:Pk_dyn}
    \begin{split}
        \frac{dP_k(t)}{dt}=\begin{cases}
        0 & t\in[t_0,t_k]\\
        AP_k(t)+P_k(t)A^\top+Q & t\in(t_{k},t_{k+1}]\\
        AP_k(t)+P_k(t)A^\top& t> t_k\;.
        \end{cases}
    \end{split}
\end{align}
Notice that substituting~\eqref{eqn:Pk_dyn} into~\eqref{eqn:P_decomp} yields the dynamics in~\eqref{eqn:prob_dyn_var}.

\subsection{Discrete-Time Feedback Covariance Dynamics}\label{subsec::feedback}
To alleviate the perturbation caused by the process noise in~\eqref{eqn:sde_lin}, a feedback control law is introduced to regulate the state deviation around the expected trajectory $\mu(t):=\E\{x(t)\}$. Based on the standard self-triggered scheme in~~\eqref{eqn:deter_ctrl_1} and~\eqref{eqn:deter_ctrl_2}, the feedback control law is defined by
\begin{align}\label{eqn:fb_ctrl}
    u(t) = \sum\limits_{k=0}^{N-1}(v_k + K(x(t_k)- \mu(t_k)))\zeta(t,t_k,t_{k+1})\;,
\end{align}
where $v_k$ is the nominal control input determined by the expected dynamics $\mu(t)$. It is noteworthy that this is a \textbf{discrete-time} linear control law written in continuous time, and it respects the self-triggered control scheme such that the control input remains constant within time interval $(t_k,t_{k+1}]$ as
\begin{align}
    u(t) = v_k + K(x(t_k)- \mu(t_k))\;,\; t\in(t_k,t_{k+1}]\;.
\end{align}
Meanwhile, as the state is accurately measured at time instance $t_0$, there is no feedback at $t_0$. 

Regarding~\eqref{eqn:prob_dyn}, the evolution of the state distribution under the control law~\eqref{eqn:fb_ctrl} is characterized by its mean and covariance, where the nominal input $v_k$ governs the mean dynamics by
\begin{align}\label{eqn:mean_dyn_ctrl}
   \frac{d\mu(t)}{dt} = A\mu(t)+B v_k\;, \forall\; t\in (t_k,t_{k+1}]\;,\;k\in\mathbb{Z}_{0}^{N-1}.
\end{align}
As the feedback part in~\eqref{eqn:fb_ctrl} reacts to the deviation from the nominal dyanmics $\mu(t)$, the covariance dynamics is therefore governed by the feedback control $K$. The following theorem gives the covariance dynamics.

\begin{theorem}\label{thm:cov_dyn}
Let the feedback control law be defined with~\eqref{eqn:fb_ctrl}, the dynamics of the covariance is given by
\begin{subequations}
\begin{align}
\frac{dP(t)}{dt} =& AP(t)+P(t)A^\top\label{eqn:cov_dyn_P}\\
&\qquad +BKP_{k,t}(t)+P_{t,k}(t)(BK)^\top+Q\;,\notag\\
\frac{dP_{t,k}(t)}{dt} =& AP_{t,k}(t)+BKP(t_k)\label{eqn:cov_dyn_Ptk}\;
\end{align}
\end{subequations}
with $P_{t,k}(t_k)=P(t_k)$. $\forall\;t\in (t_k,t_{k+1}]$ and $k\in \mathbb{Z}_{0}^{N-1}$, $P_{t,k}(t) = P_{k,t}^\top(t)$ with $$P_{t,k}(t):=\E\{\E\{(x(t)-\E\{x(t)\})(x(t_k)-\E\{x(t_k)\})^\top|\F_{k}\}\}.$$ 
\end{theorem}

A detailed proof of Theorem~\ref{thm:cov_dyn} is provided in Appendix~\ref{sect:append}. Before proceeding to the predictive control problem, we discuss the physical meaning behind Theorem~\ref{thm:cov_dyn}. $Q$ in~\eqref{eqn:cov_dyn_P} models the uncontrolled uncertainty happening during interval $(t_k,t_{k+1}]$ and $P_{t,k}$ models the stabilization effect of the feedback control law. $P_{t,k}$ in~\eqref{eqn:cov_dyn_Ptk} is the correlation between current time instance $t$ and the previous trigger moment $t_k$, which reflects the fact that the feedback control law  within $(t_k,t_{k+1}]$ only uses information up to $t_k$ to generate a constant feedback. The final piece of Theorem~\ref{thm:cov_dyn}, $P_{t,k}(t_k)=P(t_k)$, links the dynamics between $(t_{k-1},t_k]$ and $(t_k,t_{k+1}]$. In particular, as the feedback control law updates at $t_k$, $P_{t,k}$ gets reset at $t_k$ and drops the information $P_{t,k-1}(t_k)$ from the last interval.
\begin{remark}
In the first time interval $[t_0,t_1]$, we have $P_{t,0}(t_0) = 0$ and $P(t_0)=0$. Hence, the covariance dynamics in $t\in[t_0,t_1]$ is
\[       
\frac{dP(t)}{dt} = AP(t)+P(t)A^\top+Q\;,\;\;\frac{dP_{t,0}(t)}{dt} =0,
\]
which is consistent with the fact that there is no effective feedback within the first interval $[t_0,t_1]$. 
\end{remark}

\subsection{Model Predictive Control Scheme}
\label{subsec::MPC}
In this part, we show how the final stochastic MPC is formulated and the details to convert it to a solvable structure with standard stochastic MPC tricks. For the sake of compactness, the saturated resource dynamics are denoted by
\begin{align*}
g(r(t_k),\Delta_k):= \min&\{\rho \Delta_k+r(t_k)-\eta(\Delta_k),\overline{r}\}\;.
\end{align*}
In general, the nominal inputs $\{v_k\}_{k=0}^{N-1}$, the feedback control $K$, the triggering time instances $\{t_k\}_{k=1}^N$ are determined by the following problem
\begingroup\makeatletter\def\f@size{9}\check@mathfonts
\begin{subequations}\label{eqn:smpc}
\begin{align}
\min_{K,x(\cdot),v,\Delta}& \sum_{k=0}^{N-1}\int_{t_{k}}^{t_{k+1}} l(x(\tau),u(\tau))d\tau+ M(x(t_N))\\
\text{s.t.}\;\quad&\forall\,t\in (t_k,t_{k+1}],\;\forall\,k\in\{0,1,...,N-1\},\nonumber\\
 &\left\{\begin{aligned}
 \frac{d\mu(t)}{dt} &= A\mu(t)+B v_k\;,\; \\
 \frac{dP(t)}{dt} &= AP(t)+P(t)A^\top+Q\\
   +&BKP_{k,t}(t)+P_{t,k}(t)(BK)^\top,\\
\frac{dP_{t,k}(t)}{dt} &= AP_{t,k}(t)+BKP(t_k),
 \end{aligned}\right.\\
&\forall\, t\in [t_0,t_N]\;,\left\{
\begin{aligned}
x(t)& \sim\mathcal{N}(\mu(t),P(t)),\\
\mathbb{P}(x(&t)\in\mathcal{X})\geq 1-\epsilon_x,
\end{aligned}
\right.\label{eq::chance_x}\\
&\forall\, t\in \{t_k\}_{k=0}^{N-1},\left\{
\begin{aligned}
u(t_k)& {\scriptstyle\sim\mathcal{N}(K\mu(t_k),KP(t_k)K^\top)},\\
\mathbb{P}(u(&t_k)\in\mathcal{U})\geq 1-\epsilon_u,
\end{aligned}
\right.\label{eq::chance_u}\\
& \forall\,k\in\{0,1,...,N-1\}\;,\notag\\
&\left\{
\begin{aligned}
r(t_{k+1}) &= g(r(t_k),\Delta_k),\;\Delta_k\in[\underline{\Delta},\overline{\Delta}],\\
P_{t,k}(t_k)&=P(t_k),\;r(t_{k+1})\in[\underline{r},\overline{r}],
\end{aligned}
\right.
\end{align}
\end{subequations}
\endgroup
where $\epsilon_x$ and $\epsilon_u$ are the threshold that the chance constraints~\eqref{eq::chance_x} and~\eqref{eq::chance_u} are required to stay above. Notice that due to the feedback with respect to a random event, the actual input value $u(t)$ is uncertain as well. On the practical side, if the feasible $\mathcal{X}$ and $\mathcal{U}$ are polytopic, the chance constraints can be conservatively approximated by an explicit reformulation~\cite[Chapter 3]{giulioni2015stochastic}. Without loss of generality, we consider
\begin{align*}
    \mathbb{P}(H_{x,i}^\top x(t)\leq h_{x,i})\;,\;\; i\in \mathbb{Z}_{i=1}^{n_{h}}\;,
\end{align*}
where $n_{h}$ is the number of inequality constraints with respect to $x$ and $H_{x,i}\in\mathbb{R}^{n_x}$ and $h_{x,i}\in \mathbb{R}$. As $x(t)$ follows a Gaussian distribution, any of these constraints can be reformulated as
\begin{align*}
    H_{x,i}^\top \mu(t)\leq h_{x,i}-\sqrt{H_{x,i}^\top P(t)H_{x,i}}\,\mathcal{N}^{-1}(1-\epsilon_x)\;,
\end{align*}
where $\mathcal{N}^{-1}(\cdot)$ is the inverse cumulative probability distribution function, i.e.,
\[
\mathbb{P}(x\leq\mathcal{N}^{-1}(1-\epsilon_x))=1-\epsilon_x\;.
\]
\subsection{Implementation Discussion}\label{subsec::discussion}
When the problem~\eqref{eqn:smpc} is solved within a direct optimal control scheme, the integration of the ordinary differential equations can be achieved by numerical integration methods such as the Runge-Kutta-4 rule or the collocation method~\cite{levine2018handbook}. We recommend to use the collocation method, because the triggering time instances are decision variables. If Runge-Kutta is used, the integration depends on high order terms of $\{\Delta_k\}_{k=0}^{N-1}$, which results in low numerical stability. Instead, a collocation method depends linearly on $\{\Delta_k\}_{k=0}^{N-1}$ and hence is numerically more stable.

\section{Numerical Result}
\label{sec::result}
The proposed algorithm is tested on a double integrator with state $x(t)=(x_1(t),x_2(t))$, whose SDE is
\begingroup\makeatletter\def\f@size{9}\check@mathfonts
\[
dx(t) =\left(\begin{bmatrix}0 &1\\ 0&0\end{bmatrix}x(t)+\begin{bmatrix}0\\1\end{bmatrix}u(t)\right)dt + dW,\;
y(t) =
\begin{bmatrix}1\\0\end{bmatrix} x(t).
\]
\endgroup
The controller is designed to track a reference signal oscillating between $1$ and $-0.4$. Only the stage cost is considered with  
\[l(x(t),u(t))= 10(y(t)-\text{ref})^2+0.1u(t)^2\]
The parameters for the chance constraints in~\eqref{eq::chance_x} and~\eqref{eq::chance_u} are $\epsilon_x=0.01$, $\epsilon_u=0.01$. The recharging rate is $1$ with a trigger cost of $0.4$. To show the effectiveness of the proposed algorithm, we consider two different cases. In both cases the input is bounded with a chance contraint in $[-10,10]$.

In the first case, the covariance of the process noise is set to be $Q=0.01I$ and the output is bounded by $y\in[-2,1]$. In this case, the reference overlaps with the output's upper bound, and the standard deviation of the process noise is around $10\%$ the scale of the output, this case can therefore be considered as a dangerous control case. A Monte-Carlo simulation of the output responses is shown in Figure~\ref{fig:danger_output}, where the controller tries to stay close to the reference, however, as the output is upper bounded by $1$, it stays below the upper reference to ensure safety. Regarding another reference signal at $-0.4$, because it is far away from both constraints, hence the fluctuations of all the sampled experiments is centered around the desired tracking reference $-0.4$. Figure~\ref{fig:danger_resource} and Figure~\ref{fig:danger_dt} shows the responses of the resource the triggering time difference $\Delta$. When the reference is close to the bound, the controller uses the shortest triggering time confined by the resource dynamics. When the reference is further away from the bound, the resource starts to recharge. However, the resource level is lower in comparison with another case because the process noise is large and a more frequent trigger is required to guarantee the controller performance.
\definecolor{MyRed}{HTML}{e6550d}
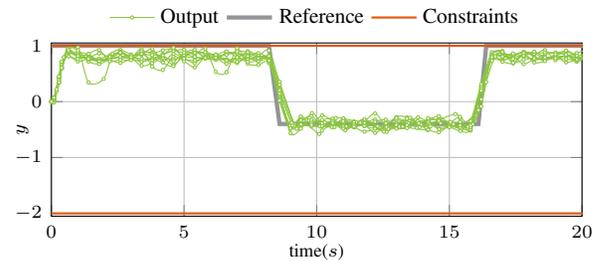
\begin{figure}[ht]
    \centering
    \begin{tikzpicture}
    \begin{axis}[xmin=0, xmax=20,
    ymin=-2.05, ymax= 1.05,
    enlargelimits=false,
    clip=true,
    grid=major,
    mark size=0.5pt,
    width=1\linewidth,
    height=0.45\linewidth,ylabel = $y$,xlabel= time($s$),
    legend style={
    	font=\footnotesize,
    	draw=none,
		at={(0.5,1.03)},
        anchor=south
    },
    legend columns=3,
    label style={font=\scriptsize},
    ylabel style={at={(axis description cs:0.12,0.5)}},
    xlabel style={at={(axis description cs:0.5,0.12)}},
    ticklabel style = {font=\scriptsize}]
    
    \pgfplotstableread{data/danger1.dat}{\dat}
    \addplot+ [thin, mark=*, mark options={fill=white, scale=1.2}, smooth,LimeGreen] table [x={t}, y={y}] {\dat};
    \addlegendentry{Output}
    \addplot+ [ultra thick,Gray,mark=none] table [x={t}, y={ref}] {\dat};
    \addlegendentry{Reference}
    \addplot+ [thick, MyRed, mark = none] table [x={t}, y={max}] {\dat};
    \addplot+ [thick, MyRed, mark = none] table [x={t}, y={min}] {\dat};
    \addlegendentry{Constraints}
    
    \pgfplotstableread{data/danger2.dat}{\dat}
    \addplot [thin, mark=*, mark options={fill=white, scale=1.2}, smooth,LimeGreen] table [x={t}, y={y}] {\dat};
    
    \pgfplotstableread{data/danger3.dat}{\dat}
    \addplot [thin, mark=*, mark options={fill=white, scale=1.2}, smooth,LimeGreen] table [x={t}, y={y}] {\dat};
    
    \pgfplotstableread{data/danger4.dat}{\dat}
    \addplot [thin, mark=*, mark options={fill=white, scale=1.2}, smooth,LimeGreen] table [x={t}, y={y}] {\dat};
    
    \pgfplotstableread{data/danger5.dat}{\dat}
    \addplot [thin, mark=*, mark options={fill=white, scale=1.2}, smooth,LimeGreen] table [x={t}, y={y}] {\dat};
    
    \pgfplotstableread{data/danger6.dat}{\dat}
    \addplot [thin, mark=*, mark options={fill=white, scale=1.2}, smooth,LimeGreen] table [x={t}, y={y}] {\dat};
    
    \pgfplotstableread{data/danger7.dat}{\dat}
    \addplot [thin, mark=*, mark options={fill=white, scale=1.2}, smooth,LimeGreen] table [x={t}, y={y}] {\dat};
    
    \pgfplotstableread{data/danger8.dat}{\dat}
    \addplot [thin, mark=*, mark options={fill=white, scale=1.2}, smooth,LimeGreen] table [x={t}, y={y}] {\dat};
    
    \pgfplotstableread{data/danger9.dat}{\dat}
    \addplot [thin, mark=*, mark options={fill=white, scale=1.2}, smooth,LimeGreen] table [x={t}, y={y}] {\dat};
    
    \pgfplotstableread{data/danger10.dat}{\dat}
    \addplot [thin, mark=*, mark options={fill=white, scale=1.2}, smooth,LimeGreen] table [x={t}, y={y}] {\dat};
    
    \end{axis}
    \end{tikzpicture}
    \caption{Output of the stochastic self-triggered MPC (Dangerous case)}
    \label{fig:danger_output}
\end{figure}

\begin{figure}[ht]
    \centering
    \begin{tikzpicture}
    \begin{axis}[xmin=0, xmax=20,
    ymin=-0.05, ymax= 1.05,
    enlargelimits=false,
    clip=true,
    grid=major,
    mark size=0.5pt,
    width=1\linewidth,
    height=0.45\linewidth,ylabel = $r$,xlabel= time($s$),
    legend style={
    	font=\footnotesize,
    	draw=none,
		at={(0.5,1.03)},
        anchor=south
    },
    ylabel style={at={(axis description cs:0.08,0.5)}},
    xlabel style={at={(axis description cs:0.5,0.12)}},
    legend columns=3,
    label style={font=\scriptsize},
    ticklabel style = {font=\scriptsize}]
    
    \pgfplotstableread{data/danger1.dat}{\dat}
    \addplot [solid, ybar,bar width=2pt, mark=none, NavyBlue] table [x={t}, y={r}] {\dat};
    \addlegendentry{Resource}
    \addplot [thick,MyRed,mark = none] table [x={t}, y={rmax}] {\dat};
    \addplot [thick,MyRed, mark = none] table [x={t}, y={rmin}] {\dat};
    \addlegendentry{Constraints}
    
    \pgfplotstableread{data/danger2.dat}{\dat}
    \addplot [ybar,bar width=2pt, mark=none, NavyBlue, solid] table [x={t}, y={r}] {\dat};
    
    \pgfplotstableread{data/danger3.dat}{\dat}
    \addplot [ybar,bar width=2pt, mark=none, NavyBlue, solid] table [x={t}, y={r}] {\dat};
    
    \pgfplotstableread{data/danger4.dat}{\dat}
    \addplot [ybar,bar width=2pt, mark=none, NavyBlue, solid] table [x={t}, y={r}] {\dat};
    
    \pgfplotstableread{data/danger5.dat}{\dat}
    \addplot [ybar,bar width=2pt, mark=none, NavyBlue, solid] table [x={t}, y={r}] {\dat};
    
    \pgfplotstableread{data/danger6.dat}{\dat}
    \addplot [ybar,bar width=2pt, mark=none, NavyBlue, solid] table [x={t}, y={r}] {\dat};
    
    \pgfplotstableread{data/danger7.dat}{\dat}
    \addplot [ybar,bar width=2pt, mark=none, NavyBlue, solid] table [x={t}, y={r}] {\dat};
    
    \pgfplotstableread{data/danger8.dat}{\dat}
    \addplot [ybar,bar width=2pt, mark=none, NavyBlue, solid] table [x={t}, y={r}] {\dat};
    
    \pgfplotstableread{data/danger9.dat}{\dat}
    \addplot [ybar,bar width=2pt, mark=none, NavyBlue, solid] table [x={t}, y={r}] {\dat};
    
    \pgfplotstableread{data/danger10.dat}{\dat}
    \addplot [ybar,bar width=2pt, mark=none, NavyBlue, solid] table [x={t}, y={r}] {\dat};
    
    \end{axis}
    \end{tikzpicture}
    \caption{Resource response of the stochastic self-triggered MPC (Dangerous case)}
    \label{fig:danger_resource}
\end{figure}
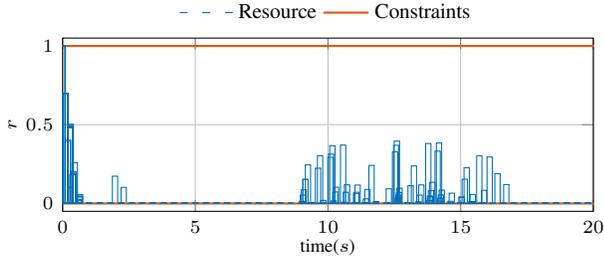

\begin{figure}[ht]
    \centering
    \begin{tikzpicture}
    \begin{axis}[xmin=0.1, xmax=20,
    ymin=0., ymax= 0.9,
    enlargelimits=false,
    clip=true,
    grid=major,
    mark size=0.5pt,
    width=1\linewidth,
    height=0.45\linewidth,ylabel = $\Delta$,xlabel= time($s$),
    legend style={
    	font=\footnotesize,
    	draw=none,
		at={(0.5,1.03)},
        anchor=south
    },
    ylabel style={at={(axis description cs:0.08,0.5)}},
    xlabel style={at={(axis description cs:0.5,0.12)}},
    legend columns=3,
    label style={font=\scriptsize},
    ticklabel style = {font=\scriptsize}]
    
    \pgfplotstableread{data/danger1.dat}{\dat}
    \addplot [scatter, only marks, mark=*, mark options={fill=white, scale=2}] table [x={t}, y={dt}] {\dat};
    \addlegendentry{Trigger time difference}
    \addplot+ [thick, MyRed, mark = none] table [x={t}, y={dtmax}] {\dat};
    \addplot+ [thick, MyRed, mark = none] table [x={t}, y={dtmin}] {\dat};
    \addlegendentry{Constraints}
    
    \pgfplotstableread{data/danger2.dat}{\dat}
    \addplot [scatter, only marks, mark=*, mark options={fill=white, scale=2}] table [x={t}, y={dt}] {\dat};
    
    \pgfplotstableread{data/danger3.dat}{\dat}
    \addplot [scatter, only marks, mark=*, mark options={fill=white, scale=2}] table [x={t}, y={dt}] {\dat};
    
    \pgfplotstableread{data/danger4.dat}{\dat}
    \addplot [scatter, only marks, mark=*, mark options={fill=white, scale=2}] table [x={t}, y={dt}] {\dat};
    
    \pgfplotstableread{data/danger5.dat}{\dat}
    \addplot [scatter, only marks, mark=*, mark options={fill=white, scale=2}] table [x={t}, y={dt}] {\dat};
    
    \pgfplotstableread{data/danger6.dat}{\dat}
    \addplot [scatter, only marks, mark=*, mark options={fill=white, scale=2}] table [x={t}, y={dt}] {\dat};
    
    \pgfplotstableread{data/danger7.dat}{\dat}
    \addplot [scatter, only marks, mark=*, mark options={fill=white, scale=2}] table [x={t}, y={dt}] {\dat};
    
    \pgfplotstableread{data/danger8.dat}{\dat}
    \addplot [scatter, only marks, mark=*, mark options={fill=white, scale=2}] table [x={t}, y={dt}] {\dat};
    
    \pgfplotstableread{data/danger9.dat}{\dat}
    \addplot [scatter, only marks, mark=*, mark options={fill=white, scale=2}] table [x={t}, y={dt}] {\dat};
    
    \pgfplotstableread{data/danger10.dat}{\dat}
    \addplot [scatter, only marks, mark=*, mark options={fill=white, scale=2}] table [x={t}, y={dt}] {\dat};
    
    \end{axis}
    \end{tikzpicture}
    \caption{Triggering time response of the stochastic self-triggered MPC (Dangerous case)}
    \label{fig:danger_dt}
\end{figure}
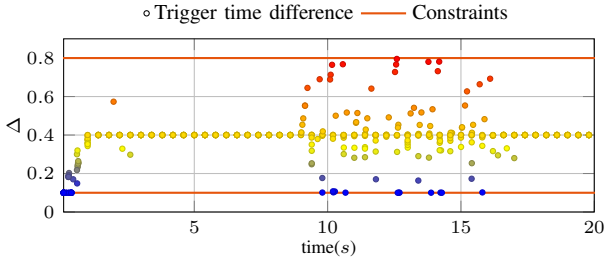

In the second case, a safer scenario is considered with smaller process noise $Q = 10^{-4}I$ and the output is bounded by $y\in[-2,1.1]$. Monte-Carlo samples of the output responses are shown in Figure~\ref{fig:safe_output}, where the output tightly tracks the reference. Meanwhile, as a stochastic control scheme, one can see that there is sampled trajectory violate the upper bound at around $1s$. To make a cleaner and more informative plot, the resource of one sampled trajectory is shown in Figure~\ref{fig:safe_resource}, where we can see that the resource tends to ramp up when the output is already around the reference and tends to decrease when the the reference signal changes. 

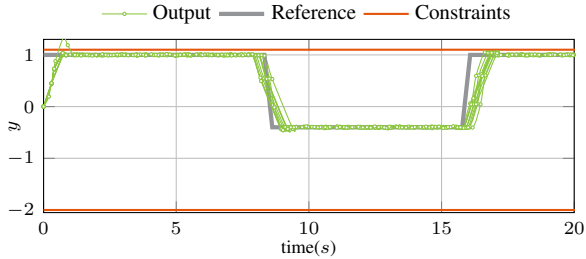
\begin{figure}[ht]
    \centering
    \begin{tikzpicture}
    \begin{axis}[xmin=0, xmax=20,
    ymin=-2.05, ymax= 1.3,
    enlargelimits=false,
    clip=true,
    grid=major,
    mark size=0.5pt,
    width=1\linewidth,
    height=0.45\linewidth,ylabel = $y$,xlabel= time($s$),
    legend style={
    	font=\footnotesize,
    	draw=none,
		at={(0.5,1.03)},
        anchor=south
    },
    ylabel style={at={(axis description cs:0.12,0.5)}},
    xlabel style={at={(axis description cs:0.5,0.12)}},
    legend columns=3,
    label style={font=\scriptsize},
    ticklabel style = {font=\scriptsize}]
    
    \pgfplotstableread{data/safe1.dat}{\dat}
    \addplot [thin, smooth,mark=*, mark options={fill=white, scale=1.2}, LimeGreen] table [x={t}, y={y}] {\dat};
    \addlegendentry{Output}
    \addplot [ultra thick,Gray,mark=none] table [x={t}, y={ref}] {\dat};
    \addlegendentry{Reference}
    \addplot [thick, MyRed, mark = none] table [x={t}, y={max}] {\dat};
    \addplot [thick, MyRed, mark = none] table [x={t}, y={min}] {\dat};
    \addlegendentry{Constraints}
    
    \pgfplotstableread{data/safe2.dat}{\dat}
    \addplot [thin, smooth,mark=*, mark options={fill=white, scale=1.2}, LimeGreen] table [x={t}, y={y}] {\dat};
    
    \pgfplotstableread{data/safe3.dat}{\dat}
    \addplot [thin, smooth,mark=*, mark options={fill=white, scale=1.2}, LimeGreen] table [x={t}, y={y}] {\dat};
    
    \pgfplotstableread{data/safe4.dat}{\dat}
    \addplot [thin, smooth,mark=*, mark options={fill=white, scale=1.2}, LimeGreen] table [x={t}, y={y}] {\dat};
    
    \pgfplotstableread{data/safe5.dat}{\dat}
    \addplot [thin, smooth,mark=*, mark options={fill=white, scale=1.2}, LimeGreen] table [x={t}, y={y}] {\dat};
    
    \pgfplotstableread{data/safe6.dat}{\dat}
    \addplot [thin, smooth,mark=*, mark options={fill=white, scale=1.2}, LimeGreen] table [x={t}, y={y}] {\dat};
    
    \pgfplotstableread{data/safe7.dat}{\dat}
    \addplot [thin, smooth,mark=*, mark options={fill=white, scale=1.2}, LimeGreen] table [x={t}, y={y}] {\dat};
    
    \pgfplotstableread{data/safe8.dat}{\dat}
    \addplot [thin, smooth,mark=*, mark options={fill=white, scale=1.2}, LimeGreen] table [x={t}, y={y}] {\dat};
    
    \pgfplotstableread{data/safe9.dat}{\dat}
    \addplot [thin, smooth,mark=*, mark options={fill=white, scale=1.2}, LimeGreen] table [x={t}, y={y}] {\dat};
    
    \pgfplotstableread{data/safe10.dat}{\dat}
    \addplot [thin, smooth,mark=*, mark options={fill=white, scale=1.2}, LimeGreen] table [x={t}, y={y}] {\dat};
    
    \end{axis}
    \end{tikzpicture}
    \caption{Output of the stochastic self-triggered MPC (Safe case)}
    \label{fig:safe_output}
\end{figure}
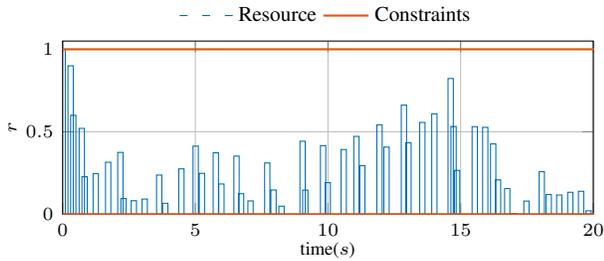
\begin{figure}[htbp!]
    \centering
    \begin{tikzpicture}
    \begin{axis}[xmin=0, xmax=20,
    ymin=0, ymax= 1.05,
    enlargelimits=false,
    clip=true,
    grid=major,
    mark size=0.5pt,
    width=1\linewidth,
    height=0.45\linewidth,ylabel = $r$,xlabel= time($s$),
    legend style={
    	font=\footnotesize,
    	draw=none,
		at={(0.5,1.03)},
        anchor=south
    },
    ylabel style={at={(axis description cs:0.08,0.5)}},
    xlabel style={at={(axis description cs:0.5,0.12)}},
    legend columns=3,
    label style={font=\scriptsize},
    ticklabel style = {font=\scriptsize}]
    
    \pgfplotstableread{data/safe2.dat}{\dat}
    \addplot+ [ybar,bar width=2pt, mark=none, NavyBlue] table [x={t}, y={r}] {\dat};
    \addlegendentry{Resource}
    \addplot+ [thick, MyRed, mark = none] table [x={t}, y={rmax}] {\dat};
    \addplot+ [thick, MyRed, mark = none] table [x={t}, y={rmin}] {\dat};
    \addlegendentry{Constraints}

    \end{axis}
    \end{tikzpicture}
    \caption{Resource response of the stochastic self-triggered MPC (Safe case)}
    \label{fig:safe_resource}
\end{figure}

\section{Conclusion}
\label{sec::concl}
This work proposes a novel resource-aware stochastic self-triggered MPC, which generalizes resource-aware self-triggered MPC to an uncertain environment. The discrete time covariance dynamics of a discrete-time feedback control law is derived to accommodate a continuous time uncertain disturbance. This discrete feedback scheme is intentionally designed to be compatible with a self-triggered control scheme. Finally, the proposed scheme is validated through a numerical example.

{\tiny
\bibliographystyle{abbrv}
\bibliography{ref.bib}}

\begin{thebibliography}{10}

\bibitem{aydiner2015robust}
E.~Aydiner, F.~D. Brunner, W.~Heemels, et~al.
\newblock Robust self-triggered model predictive control for constrained
  discrete-time lti systems based on homothetic tubes.
\newblock In {\em 2015 European Control Conference (ECC)}, pages 1587--1593,
  2015.

\bibitem{ben2009robust}
A.~Ben-Tal, L.~El~Ghaoui, and A.~Nemirovski.
\newblock {\em Robust optimization}.
\newblock Princeton university press, 2009.

\bibitem{berglind2012self}
J.~B. Berglind, T.~Gommans, and W.~Heemels.
\newblock Self-triggered mpc for constrained linear systems and quadratic
  costs.
\newblock {\em IFAC Proceedings Volumes}, 45(17):342--348, 2012.

\bibitem{bernardini2012energy}
D.~Bernardini and A.~Bemporad.
\newblock Energy-aware robust model predictive control based on noisy wireless
  sensors.
\newblock {\em Automatica}, 48(1):36--44, 2012.

\bibitem{bock1984multiple}
H.~G. Bock and K.-J. Plitt.
\newblock A multiple shooting algorithm for direct solution of optimal control
  problems.
\newblock {\em IFAC Proceedings Volumes}, 17(2):1603--1608, 1984.

\bibitem{brunner2016robust}
F.~D. Brunner, M.~Heemels, and F.~Allg{\"o}wer.
\newblock Robust self-triggered mpc for constrained linear systems: A
  tube-based approach.
\newblock {\em Automatica}, 72:73--83, 2016.

\bibitem{farina2012tube}
M.~Farina and R.~Scattolini.
\newblock Tube-based robust sampled-data mpc for linear continuous-time
  systems.
\newblock {\em Automatica}, 48(7):1473--1476, 2012.

\bibitem{giulioni2015stochastic}
L.~Giulioni.
\newblock {\em Stochastic model predictive control with application to
  distributed control systems}.
\newblock PhD thesis, Italy, 2015.

\bibitem{heemels2012introduction}
W.~Heemels, K.~H. Johansson, and P.~Tabuada.
\newblock An introduction to event-triggered and self-triggered control.
\newblock In {\em 2012 IEEE 51st IEEE Conference on Decision and Control
  (CDC)}, pages 3270--3285, 2012.

\bibitem{henriksson2012self}
E.~Henriksson, D.~E. Quevedo, H.~Sandberg, and K.~H. Johansson.
\newblock Self-triggered model predictive control for network scheduling and
  control.
\newblock {\em IFAC Proceedings Volumes}, 45(15):432--438, 2012.

\bibitem{krishnan2013nonlinear}
V.~Krishnan.
\newblock {\em Nonlinear filtering and smoothing: An introduction to
  martingales, stochastic integrals and estimation}.
\newblock Courier Corporation, 2013.

\bibitem{le2016brownian}
J.-F. Le~Gall.
\newblock {\em Brownian motion, martingales, and stochastic calculus}.
\newblock Springer, 2016.

\bibitem{levine2018handbook}
W.~S. Levine, L.~Gr{\"u}ne, et~al.
\newblock Handbook of model predictive control.
\newblock 2018.

\bibitem{li2014event}
H.~Li and Y.~Shi.
\newblock Event-triggered robust model predictive control of continuous-time
  nonlinear systems.
\newblock {\em Automatica}, 50(5):1507--1513, 2014.

\bibitem{lian2020resource}
Y.~Lian, S.~Wildhagen, Y.~Jiang, B.~Houska, F.~Allg{\"o}wer, and C.~N. Jones.
\newblock Resource-aware asynchronous multi-agent coordination via
  self-triggered mpc.
\newblock In {\em 2020 59th IEEE Conference on Decision and Control (CDC)},
  pages 685--690, 2020.

\bibitem{liu2018robust}
C.~Liu, H.~Li, J.~Gao, and D.~Xu.
\newblock Robust self-triggered min--max model predictive control for
  discrete-time nonlinear systems.
\newblock {\em Automatica}, 89:333--339, 2018.

\bibitem{wildhagen2019resource}
S.~Wildhagen, C.~N. Jones, and F.~Allg{\"o}wer.
\newblock A resource-aware approach to self-triggered model predictive control.
\newblock {\em arXiv preprint arXiv:1911.10799}, 2019.

\end{thebibliography}
\small
\section{Appendix}\label{sect:append}
\subsection*{Appendix to Section~\ref{subsec::decomp}}
We denote the natural filtration generated by a Wiener process $W$ as $\F_t$ with a trivial $\F_0 = \{0,\Sigma\}$. Here, $\Sigma$ is the whole event space. We remind the definition $\F_{k,k+1}:=\sigma(\F_{k+1}\backslash\F_{k})$\footnote{This definition holds for any Lévy processes, but not necessarily for any filtration.}. The proof of Lemma~\ref{lem:decomp_sigma} is\\
\textbf{Proof of Lemma~\ref{lem:decomp_sigma}}
According to the independence property of a Wiener process, $\F_{i,i+1}\perp\F_{j,j+1}\;$ holds for all $i\neq j$, therefore we have $\sigma(\cup_{i=0}^{N-1}\F_{i,i+1})\subset \F_{N}$. Then, we show the equality holds by contradiction. If $\F_{N} \neq \sigma(\cup_{i=0}^{N-1}\F_{i,i+1})$, by definition of $\F_{i,i+1}$, there exists $i\in\mathbb{Z}_{0}^{N-1}$ such that $\F_{t_i}\neq\F_{t_{i}^+}$\footnote{$t_i^+:=t>t_i,\;,t\rightarrow t_i$, $\F_{t_{i}^+}:=\sigma(\cap_{t>t_i}\F_t)$}, which violates the continuity of a Wiener process~\cite{le2016brownian}. Hence, the proof concludes.
%\hfill$\blacksquare$
\begin{remark}
This lemma holds for any càdlàg Lévy process, which is practical for real-world applications as all the analysis is established from the current time step or, in particular, the sigma fields accumulated up to the current time instance.
\end{remark}
\begin{remark}
Equation~\eqref{eqn:P_decomp} holds due to product topology given by the Lemma~\ref{lem:decomp_sigma}, which reflects the fact that the conditional covariance matrix $P_k(t):=\E(P(t)|\F_{k,k+1})$ is a projection onto the $\mathcal{L}^2$ space of the progressively measurable process on $\F_{k,k+1}$.
\end{remark}
\subsection*{Appendix to Section~\ref{subsec::feedback}}
Before proceeding to the proof, the left-open right-closed time intervals used in this paper, such as Theorem~\ref{thm:cov_dyn} and Equation~\eqref{eqn:Pk_dyn}, stresses the continuity of the considered stochastic process. As the dynamics are Lipschitz, a bounded step change in the control input will not change the solution and, in particular, it ensures a unique strong solution~\cite{le2016brownian}.

To prove Theorem~\ref{thm:cov_dyn}, we introduce the It\^o's Lemma~\cite{le2016brownian}.
\begin{lemma}[It\^o's Lemma]\label{lem::ito}
For a given drift-diffusion process $dx = adt+bdW$, if function $f(\cdot)$ is twice-differentiable, I\^to's formula holds as 
\[
df = \left(\frac{\partial f}{\partial t}+a\frac{\partial f}{\partial x}+\frac{b^2}{2}\frac{\partial^2 f}{\partial x^2}\right) dt + b\frac{\partial f}{\partial x}dW\;.
\]
\end{lemma}
This lemma quantifies the function evolution driven by an SDE.
\textbf{Proof of Theorem~\ref{thm:cov_dyn}.}
In this proof, we will first show the continuous time dynamics of the decomposed matrices $\{P_k(\cdot)\}_{k=0}^{N-1}$ , which is then used to reconstruct the continuous time dynamics of $P(t)$ based on Lemma~\ref{lem:decomp_sigma}.

First, we show the evolution of the covariance $P_k(t)$. Conditioning on the sigma field $\F_{k,k+1}$, the control law is
\begin{align}\label{eqn:cond_ctrl}
&u(t\,|\,\F_{k,k+1}) =\\\notag 
&\begin{cases}
v_i  &t\in (t_i,t_{i+1}],i\in\mathbb{Z}_0^k,\\[0.16cm]
v_i+K\E\{x(t)-\mu(t)|\F_{k,k+1}\} & t\in(t_i,t_{i+1}], i\in\mathbb{Z}_{k+1}^{N-1}.
\end{cases}
\end{align}
Notice that under a predictive control scheme, $\{v_i\}_{i=0}^{N-1}$ are determined at $t_0$, hence $\{v_i\}_{i=0}^{N-1}$ are $\F_0$ measurable and furthermore $\F_{k,k+1}$ is measurable. Before $t_{k+1}$, none of the triggers can generate feedback with respect to the events in $\F_{k,k+1}$ because $\F_{k,k+1}$ happens later than $\{t_i\}_{i=0}^{k}$. These facts conclude the conditional control law in~\eqref{eqn:cond_ctrl}. Based on the system dynamics~\eqref{eqn:sde_lin}, the mean dynamics~\eqref{eqn:mean_dyn_ctrl} and the conditional control inputs~\eqref{eqn:cond_ctrl}, the SDE of the conditional deviation dynamics of $x(t)-\mu(t)$ is
\begin{align}
    &\mathbb{E}\{d(x(t)-\mu(t))|\F_{k,k+1}\}=\\\notag
    &\begin{cases}
    0& t\in [t_0,t_k],\\[0.2cm]
    A(\mathbb{E}\{x(t)-\mu(t)|\F_{k,k+1}\})dt+dW&t\in(t_k,t_{k+1}],\\[0.2cm]
    \left[A(\mathbb{E}\{x(t)-\mu(t)|\F_{k,k+1}\})+B\right.& t\in(t_i,t_{i+1}],\\
    \;\;\left.\cdot K(\mathbb{E}\{d(x(t_i)-\mu(t_i))|\F_{k,k+1}\})\right]dt&i\in\mathbb{Z}_{k+1}^{N-1}.
    \end{cases}
\end{align}
More specifically, this dynamics means that the stochastic events within interval $(t_k,t_{k+1}]$ do not generate any feedback before $t_{k+1}$ and the deviation evolves in an open-loop form. After $t_{k+1}$, no new $\F_{k,k+1}$-measurable events can happen anymore and the feedback control law comes into effect.

As \begingroup\makeatletter\def\f@size{9}\check@mathfonts$P_k(t) = \E\{\E\{(x(t)-\mu(t))(x(t)-\mu(t))^\top|\F_{k,k+1}\}\}$\endgroup, we can apply Itô's Lemma (Lemma~\ref{lem::ito}) to the deviation dynamics. As a result, we have
\begin{align*}
    \forall\,t\in[t_0,t_k]\;,\;\;\frac{dP_k(t)}{dt} = 0\;.
\end{align*}
And for all $t\in(t_k,t_{k+1}]$, we have 
\begin{align*}
    dP_k(t) \;=\; &(AP_k(t)+P_k(t)A^\top+Q)dt\\
    &+\underbrace{\E\{\E\{x(t)-\mu(t)|\F_{k,k+1}\}dW^\top\}}_{(a)}\\
    &+\underbrace{\E[dW\E(x(t)-\mu(t)|\F_{k,k+1})^\top]}_{(b)}\;,
\end{align*}
where $(a)=0$ as $\E\{\E\{x(t)-\mu(t)|\F_{k,k+1}\}\}=0$, and similarly, $(b)=0$. We thus, conclude
\begin{align*}
    \frac{dP_k(t)}{dt} = AP_k(t)+P_k(t)A^\top+Q\;,\forall\, t\in(t_k,t_{k+1}]\;,
\end{align*}
which shares a form similar to~\eqref{eqn:prob_dyn_var}.

The last piece is the intervals in which the feedback control law takes effect. Without loss of generality, we consider one interval $(t_i,t_{i+1}]$ with $i\in\mathbb{Z}_{k+1}^{N-1}$, where we have 
\begin{align*}
    \frac{dP_k(t)}{dt} =A P_k(t)+ P_k(t)A^\top
    +BK P_{i,t,k}(t)+ P_{t,i,k}(t)(BK)^\top\;
\end{align*}
with $P_{t,i,k}(t)=P_{i,t,k}(t)^\top:=$
\begin{align*}
\E\{\E\{(x(t)-\mu(t))(x(t_i)-\mu(t_i)^\top\}|\F_{k,k+1}\}\;.
\end{align*}
Applying Itô's Lemma again, we have
\begin{align*}
    \frac{dP_{t,i,k}}{dt}=AP_{t,i,k}+BKP_{t_i,i,k}(t_i)=AP_{t,i,k}+BKP_{k}(t_i)\;,
\end{align*}
where the second equality holds by definition. 

As the conditional covariance dynamics is available, we are ready to conclude the general dynamics. Considering interval $(t_i,t_{i+1}]$, we have following facts:
\begin{enumerate}
    \item $\forall\,k\geq i+1$, we have $P_k(t)= 0$.
    \item Because the feedback is not active for the sigma-fields $\F_{k,k+1},\;\forall\; k\geq i$, we have $P_{t,i,k}= ,\;\forall\;k\geq i0$.
\end{enumerate}
Based on the previous derivation, we have
\begingroup\makeatletter\def\f@size{8.5}\check@mathfonts
\begin{align*}
    \frac{dP(t)}{dt} \overset{(a)}{=}\sum\limits_{k=0}^{N-1}\frac{dP_k(t)}{dt}
    =\underbrace{\sum\limits_{k=0}^{i-1}\frac{dP_k(t)}{dt}}_{\circled{1}}+\underbrace{\frac{dP_i(t)}{dt}}_{\circled{2}}+\underbrace{\sum\limits_{k=i+1}^{N-1}\frac{dP_k(t)}{dt}}_{\circled{3}}\\
    = \underbrace{\sum_{k=0}^{i}(AP_i(t)+P_i(t)A^\top)}_{(b)}+Q + \underbrace{\sum_{k=0}^{i-1}BK P_{k,t,i}(t)+ P_{t,k,i}(BK)^\top}_{(c)}\;,
\end{align*}\endgroup
where $(a)$ holds by Lemma~\ref{lem:decomp_sigma}. \circled{1} corresponds to the components whose feedback is active, \circled{2} incorporates the stochastic event happening in the current interval $(t_i,t_{i+1}]$, while \circled{3} are the future stochastic events which have no effect yet. The first aforementioned fact allows the reformulation of $(b)$ as
\begin{align*}
    (b) = \sum_{k=0}^{N-1}(AP_k(t)+P_k(t)A^\top)\;.
\end{align*}
Similarly, the second aforementioned fact reformulates $(c)$ as
\begin{align*}
    (c)=\sum_{k=0}^{N-1}BK P_{i,t,k}(t)+ P_{t,i,k}(t)(BK)^\top\;.
\end{align*}
Hence, by equation~\eqref{eqn:P_decomp}, we conclude
\begingroup\makeatletter\def\f@size{8.8}\check@mathfonts
\[
\frac{dP(t)}{dt} =AP(t)+P(t)A^\top+BKP_{k,t}(t)+P_{t,k}(t)(BK)^\top+Q\;.
\]
\endgroup
In a similar approach, we have 
\begin{align*}
    \frac{dP_{t,k}(t)}{dt} = AP_{t,k}(t)+BKP(t_k)\;,
\end{align*}
which concludes the proof. %\hfill$\blacksquare$

\end{document}